\documentstyle[epsfig]{aipproc}
\begin{document} \title{Detection of Gravitational Waves from Gravitationally Lensed Systems} \author{T. Wickramasinghe$^*$ and M. Benacquista$^{\dagger}$}
\address{$^*$Dept. of Physics, The College of New Jersey, Ewing, NJ 08628\\
$^{\dagger}$Dept. of Sciences, Montana State University-Billings, Billings, MT 59101}
\maketitle
\begin{abstract}
It is accepted that quasars are powered by supermassive black holes (SMBH) with masses in the range $10^6 - 10^9 M_{\sun}$ in their cores. Occasionally, compact stars can plunge into SMBH. In addition, there may be a number of such compact objects circling the central SMBH in any given quasar. Both of these processes are known to emit gravitational waves. LISA has the right sensitivity to detect these waves. We show that gravitational lenses amplify the amplitudes of these gravitational waves just as they amplify the observed light of quasars. Given the geometry of the lensing configuration, this amplification can be as large as a factor of 2 to 10, allowing the waves to be above the detection threshold of LISA. We also show that waves from lensed quasars arrive with time delays which are much larger than the coherence time of the gravitational waves, making interference effects negligible. Thus, a simple geometrical optics application leads to the lensing theory of gravitational waves. In this context, we an alyze and show in this preliminary analysis that there is an enhancement of the amplitudes of gravitational radiation coming from observed lensed quasars.
\end{abstract}

It is now well accepted that the energy source of a quasar is a supermassive black hole (SMBH)\cite{hoyle}. The mass of the SMBH is of the order of $10^6 - 10^9 M_{\sun}$ \cite{rees}. In this mechanism of power generation, a compact star circling the SMBH plunges into it. Both these processes, encircling and plunging, are known to emit copious amounts of gravitational radiation \cite{thorne,cutler}. These waves can in principle undergo gravitational lensing just as electromagnetic radiation does. We shall analyze the lensing of such waves.

A compact star plunging into the SMBH at the core of a quasar produces a burst of gravitational radiation which can be approximated as having gravitational strain $h$ and frequency $f$. These two quantities are given by \cite{haehnelt,shibata}:
\begin{eqnarray}
h &\simeq& 10^{-21}\left(\frac{10 Mpc}{R}\right)\left(\frac{m}{1 M_{\sun}}\right)\label{strain}\\
f &\simeq& (1.3 \times 10^{-2} Hz)\left(\frac{10^6 M_{\sun}}{M}\right)
\label{freq}
\end{eqnarray}
where $R$ is the distance to the system, $m$ is the mass of the compact object, and $M$ is the mass of the SMBH. LISA has enough sensitivity to detect gravitational waves with amplitudes $h \sim 10^{-20} - 10^{-21}$ and frequencies $f \sim 10^{-1} - 10^{-3} Hz$ \cite{shibata}.

There are a number of gravitationally lensed quasars within $z \sim 1$ \cite{kembhavi}. The lensing object usually is a normal galaxy having a mass of the order of $10^{11} M_{\sun}$. The radius of curvature of the spacetime in the proximity of the lensing galaxy is $R_{\rm curvature} \sim GM_l/c^2$, where $M_l$ is the mass of the lensing galaxy. Eq.~\ref{freq} shows that $\lambda_{GW} \sim 10^{7} km$, while $R_{\rm curvature} \sim 10^{11} km$. Thus $\lambda_{GW} / R_{\rm curvature} << 1$ and we can use a simple geometrical optics approximation to analyze lensing of gravitational waves coming from lensed quasars. Gravitational waves of a lensed quasar should bend in the proximity of the lens, just as light coming from the quasar. This lensing effect will produce multiple images of the quasar due to area distortion near the lens.

\begin{figure}[h]
\centerline{\epsfig{file=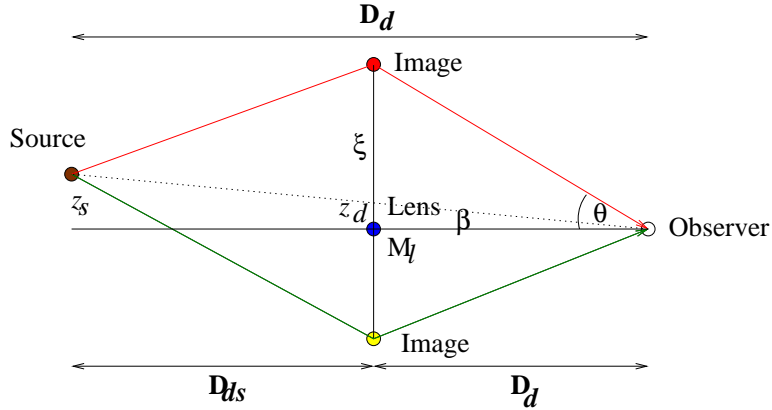,width=4in}}
\caption{The general gravitational lensing geometry. Lens - Observer is the optic axis. $\beta$ and $\theta$ are the source and image positions respectively. $z_s$ and $z_d$ are the redshifts of the source and the lens.}
\label{geometry}
\end{figure}

The distances in Fig.~\ref{geometry} are Dyer and Roder angular diameter distance \cite{dyer}. If the universe is filled with a smooth distribution of matter, then the angular diameter distance measured from a point at redshift $z_1$ to a point at $z_2$ is given by:
\begin{equation}
D_A(z_1,z_2) = \frac{c}{H_0}r(z_1,z_2) = 2\left[\frac{1}{(1+z_2)\sqrt{1+z_1}}-\frac{1}{(1+z_2)^{3/2}}\right].
\end{equation}
The luminosity distance is given by: $D_l = D_A(1+z)^2$. In Eq.~\ref{strain}, we must replace $R$ by $D_l$. An Einstein ring forms when the perfect alignment is achieved. The radius of this ring is $\xi_0 = \sqrt{4GM_lD_dD_{ds}/(c^2 D_s)}$ \cite{schneider}, and its angular size is $2\alpha_0$. We now define dimensionless quantities $y = \beta/\alpha_0$ and $x = \theta/\alpha_0$ so that $y$ and $x$ are the dimensionless positions of the source and image, respectively.
With $c/H_0 = 2997.9/h_0 \; \mbox{Mpc}$, we derive the following relationships ($\alpha_0'' = \alpha_0$ in arcseconds):
\begin{eqnarray}
\alpha_0'' &=& 1.65\times10^{-6}\sqrt{h_0}\sqrt{\frac{M_l}{M_{\sun}}}\psi(z_d,z_s)\label{alphaeq}\\ \psi(z_d,z_s) &=& \sqrt{\frac{r(z_d,z_s)}{r(0,z_s)r(0,z_d)}}\label{psieq}\\
\mu &=& \frac{y^2+2}{y\sqrt{y^2+4}}\label{mueq}\\
c\Delta t &=& 1/\vert d\beta/d\theta\vert =\frac{4GM_l}{c^2}(1+z_d)\left[\frac{1/2}y\sqrt{y^2+4} + \ln{\frac{\sqrt{y^2+4}+y}{\sqrt{y^2+4}-y}}\right]\label{cdt}
\end{eqnarray}
Equations~\ref{mueq} and~\ref{cdt} are well-known equations in gravitational lensing theory \cite{schneider}. Eq~\ref{mueq} gives the total amplification of images. Eq.~\ref{cdt} shows that there is a time delay $\Delta t$ between any two images.

The equations~\ref{alphaeq} through ~\ref{cdt} can be used to estimate the amplification of the gravitational waves coming from a lensed quasar. Eq.~\ref{cdt} shows that the time delay between any two images of a realistic gravitational lensing configuration of a quasar is many days. Comparison with $f_{GW}$ shows that interference effects can safely be neglected. LISA's angular resolution scales as the inverse of the signal to noise and is a few square degrees. The separate images of a quasar are usually separated by a few arcseconds. Therefore, it is hopeless to imagine that LISA will be able to resolve images. However, it will see a composite, amplified image. The gravitational wave strain coming from this object must then be $\sim \mu h$.

For any practical astrophysical situation, $\psi(z_d,z_s) < 2$. Note that $y < 1$ for $\beta < \alpha_0$ when the source is within the Einstein ring of the lens. We note from Eq.~\ref{mueq} that the amplification in this case is large. It is clear also that substantially small source positions $y$ can be expected in real astrophysical situations which implies that gravity waves coming from lensed quasars are typically amplified by a factor of 2 - 10. We see from Eq.~\ref{strain} that the sensitivity of LISA is sufficient to detect gravitational waves emanating from a gravitationally lensed quasar system as far distant as 100 Mpc for an amplification as large as 10. We stress the fact that our analysis is still preliminary.

\end{document}